\pdfoutput=1
\documentclass[apj]{emulateapj}
\usepackage[bookmarks=true,colorlinks=true,citecolor=blue,linkcolor=magenta,urlcolor=cyan]{hyperref}

\usepackage{natbib}
\usepackage{amsmath}
\usepackage{amssymb}
\usepackage{subfigure}
\usepackage{url}
\usepackage{CJK}

\newcommand{\rmd}{ {\ \mathrm d} }
\renewcommand{\vec}[1]{ {\mathbf #1} }

\newcommand{\curl}{ {\bf \nabla} \times}

\newcommand{\Eq}{{Equation}}

\newcommand{\Fig}{{Figure}}

\newcommand{\divB}{\nabla\cdot\mathbf{B}}
\newcommand{\crlB}{\nabla\times\mathbf{B}}

\newcommand{\SDO}{{\it SDO}}

\slugcomment{Accepted for publication in ApJ}

\shorttitle{Magnetic Field Extrapolation with \SDO/HMI Data}
\shortauthors{Jiang et al.}

\begin{document}
\begin{CJK*}{UTF8}{gbsn}


\title{Extrapolation of the Solar Coronal Magnetic Field from
  \SDO/HMI Magnetogram by a CESE--MHD--NLFFF Code}


\author{
  Chaowei Jiang (江朝伟)\altaffilmark{1},
  Xueshang Feng (冯学尚)\altaffilmark{1}}
\email{cwjiang@spaceweather.ac.cn, fengx@spaceweather.ac.cn}


\altaffiltext{1}{SIGMA Weather Group, State Key Laboratory for Space
  Weather, Center for Space Science and Applied Research, Chinese
  Academy of Sciences, Beijing 100190}


\begin{abstract}
  Due to the absence of direct measurement, the magnetic field in
  the solar corona is usually extrapolated from the photosphere in
  numerical way. At the moment, the nonlinear force-free field (NLFFF)
  model dominates the physical models for field
  extrapolation in the low corona. Recently we have developed a
  new NLFFF model with MHD relaxation to reconstruct the
  coronal magnetic field. This method is based on
  CESE--MHD model with the conservation-element/solution-element
  (CESE) spacetime scheme. In this paper, we report the application of
  the CESE--MHD--NLFFF code to \SDO/HMI data with magnetograms sampled
  for two active regions (ARs), NOAA AR 11158 and 11283, both of which
  were very
  non-potential, producing X-class flares and eruptions. The raw magnetograms are preprocessed to remove the
  force and then inputted into the extrapolation code. Qualitative
  comparison of the results with the \SDO/AIA images shows that our
  code can reconstruct magnetic field lines resembling the
  EUV-observed coronal loops. Most important structures of the active
  regions are reproduced excellently, like the highly-sheared field
  lines that suspend filaments in AR~11158 and twisted flux rope
  which corresponds to a sigmoid in AR~11283. Quantitative assess of
  the results shows that the force-free constraint is fulfilled very
  well in the strong-field regions but apparently not that well in the
  weak-field regions because of data noise and numerical
  errors in the small currents.
\end{abstract}






\section{Introduction}
\label{sec:intr}

Magnetic field extrapolation is an important tool to study the
three-dimensional (3D) solar coronal magnetic field, which is still
difficult to measure directly \citep{Sakurai1989, Aly1989, Amari1997,
  McClymont1997, Wiegelmann2008, DeRosa2009}. The models being used
most popularly for field extrapolation are the potential field model,
the linear force-free field model, and the nonlinear force-free field
(NLFFF) model. These models are all based on the same assumption that
the Lorentz force is self-balancing in the corona, but adopt different
simplifications of the current distribution. Among these models, The
NLFFF is the most used one for characterizing magnetic field in the
low corona, where there is significant and localized electric current,
especially in the active regions (ARs).

But, to directly solve the general NLFFF equation
\begin{equation}
  \label{eq:ff}
  (\crlB) \times \vec B = \vec 0,\ \
  \divB = 0
\end{equation}
is really difficult. As is known, the system is nonlinear
intrinsically and even the existence and uniqueness of a solution for
a given boundary condition are not proved theoretically; solutions
have rarely been found in closed analytic form
\citep[e.g.,][]{Low1990} and in most cases people can only resort to
numerical method using computer \citep[many numerical codes have been
developed in the past decades, e.g.,][one may refer to a recent living
review by \citet{Wiegelmann2012solar}]{Wu1990, Roumeliotis1996,
  Amari1999, Wheatland2000, Yan2000, Wiegelmann2004, Valori2007,
  Inoue2011}. Moreover the observation can only provide a bottom
boundary of data, and even worse, on the photosphere the field is
forced significantly by the dense plasma and thus conflicts with the
fundamental force-free assumption. Besides the noise in the observation,
measurement error and instrumental uncertainty (e.g., the well
known $180^{\circ}$ ambiguity of the transverse fields) are all
rather unfavorable for practical computation. Thus the
observed magnetogram usually needs to be preprocessed to remove the
force and noise for providing a better input
\citep{Wiegelmann2006b}. Anyhow, at present one can hardly seek an
exact force-free solution with the observation information fully
satisfied. The best we can do is to find a good balance between the
force-free constraint and deviation from the real observation, i.e.,
to seek an approximately force-free solution that matches the
photospheric field measurements as well as possible.

Recently we have developed a new extrapolation code called
CESE--MHD--NLFFF\footnote{We initially planned to develop a full MHD
 model for computing both the static non-potential field and dynamic
 evolution of ARs \citep[i.e., the CESE--MHD model,][]{Jiang2011}.
 But we find that the MHD solver is rather slow to construct the static field, although
 success has been reported in \citet{Jiang2012c}.
 We thus move to develop a NLFFF version of this model for a faster convergence speed,
 while the full MHD solver is more suitable for simulating transient
 events like the eruptions.} \citep{Jiang2011,Jiang2012apj}, which is based on
magnetohydrodynamics (MHD) relaxation method and an advanced numerical
scheme, the spacetime conservation-element/solution-element (CESE)
method, for faster convergence and better accuracy over the
available codes. The good performance and high accuracy of the code
have been demonstrated through critical comparisons with previous
joint studies by \citet{Schrijver2006} and \citet{Metcalf2008}, in
which various NLFFF codes are assessed based on several NLFFF
benchmark tests. We have also successfully extended the code to
application in spherical geometry and seamless full-sphere
extrapolation for the global corona \citep{Jiang2012apj1}.

In this paper we report the application of the CESE--MHD--NLFFF code to
real solar data, i.e., the presently released \SDO/HMI
magnetograms. To deal with the real observation data, we also have
developed a new preprocessing method to remove the force in the raw
magnetogram \citep{Jiang2012SoPsubmit}. Extra advancements are made to the
original code to further enhance the ability of handling the
high-resolution but noisy data. Magnetograms of two ARs, AR~11158 and
AR~11283 are sampled for our tests of extrapolation. The results are
carefully assessed both qualitatively and quantitatively. We show that
our code can recover magnetic field lines resembling the plasma loops
seen in the \SDO/AIA images, and reproduce most important structures of the ARs
remarkably well like the highly-sheared field lines that suspend
filaments in AR~11158 and twisted flux rope that corresponds to a
sigmoid in AR~11283.

The remainder of the paper is organized as follows. We first describe
briefly the CESE--MHD--NLFFF code along with its improvements in
Section~\ref{sec:code}. The magnetogram from \SDO/HMI and the
preprocessed result of the raw data are given in
Section~\ref{sec:data}. We then present the extrapolation results for
these data including both the raw and preprocessed magnetogram in
Section~\ref{sec:resu}. Finally, we draw conclusions and give some
outlooks for future work in Section~\ref{sec:colu}.

\section{The CESE--MHD--NLFFF Code}
\label{sec:code}

The basic idea of using the MHD relaxation approach to solve the
force-free field is to use some kind of fictitious dissipation to
drive the MHD system to an equilibrium in which all the forces can be
neglected comparing with the Lorentz force while the boundary magnetogram
is satisfied. In this way the Lorentz force should be self-balancing
and the field can be regarded as the target force-free solution. We
solve the magneto-frictional model equations in the magnetic splitting
form as
\begin{eqnarray}
  \label{eq:main_equ}
  \frac{\partial\rho\mathbf{v}}{\partial t} =
  (\crlB_{1})\times\mathbf{B}-(\divB_{1})\vec B-\nu\rho\mathbf{v},
  \nonumber\\
  \frac{\partial\mathbf{B_{1}}}{\partial t} =
  \nabla\times(\mathbf{v}\times\mathbf{B})
  +\nabla(\mu\divB_{1})
  -\mathbf{v}\divB_{1},
  \nonumber\\
  \rho=|\vec B|^{2}+\rho_{0},\
  \mathbf{B}=\mathbf{B}_{0}+\mathbf{B}_{1}.
\end{eqnarray}
$\vec B$ is the target force-free field to be solved, $\vec
B_{0}$ is the potential field matching the normal component of the
magnetogram, and $\vec B_{1}$ is the deviation between $\vec B$ and
$\vec B_{0}$. $\nu$ is the frictional coefficient and $\mu$ is the
numerical diffusive speed of the magnetic monopole. The value of them
are respectively given by $\nu=1/(5\Delta t)$ and $\mu=0.4(\Delta
x)^{2}/\Delta t$ in the code, according to the time step $\Delta t$ and
local grid size $\Delta x$. Many advantages can be gained by solving
such form of above equations \citep{Jiang2012apj,Jiang2012apj1}.

The above equation system~(\ref{eq:main_equ}) is solved by our
CESE--MHD scheme \citep{Jiang2010}. In principle we can use any
available MHD code to solve this set of equations, since it is a
subset of the full MHD system. Taking consideration of the
computational efficiency and accuracy, we prefer to utilize modern
advanced MHD codes. However, most of the modern MHD codes are based on
theory of characteristic decomposition of a hyperbolic system,
thus are not suitable for equation (2), which is not a hyperbolic
system. The CESE scheme is a new method free of characteristic
decomposition and is very suitable for the equation form
here. Furthermore, the CESE-MHD code has been extensively used
in solar physics, e.g., the data-driven evolution
modeling of AR \citep{Jiang2012c}, the global corona
\citep{Feng2012apj} and the interplanetary solar wind
\citep{Feng2012SoPh,Yang2012JGR}.

To adapt for the application to real solar data, we have made extra
improvements to the previous version of CESE--MHD--NLFFF. The first
improvement is made to enhance the ability of handling noisy data in
the magnetograms. In the noisy weak-field regions of
magnetograms (where the signal-to-noise ratio is small, say $|\vec
B|\le 100$~G), the term $(\curl\vec B)\times \vec B/|\vec B|^{2}$ could
be very large due to numerical gradients of the random noise, thus the
velocity $\vec v$ is prone to be accelerated to extremely high, which
can severely restrict the time step and slow the relaxation process of
the entire system, even making the computation unmanageable. To deal
with this difficulty, the pseudo-plasma density $\rho$ is designed
with $\rho_{0} = B_{\min}^{2}\exp(-z/H_{\rm m})$, where $z$ is the
height from the bottom surface and $B_{\min}=100$~G, $H_{\rm
  m}=5$~pixel. In this way, the velocity (near the bottom magnetogram)
in the weak-field regions can be reduced significantly, while it is barely affected
in the strong-field regions.

\begin{figure}[htbp]
  \centering
  \includegraphics[width=0.40\textwidth]{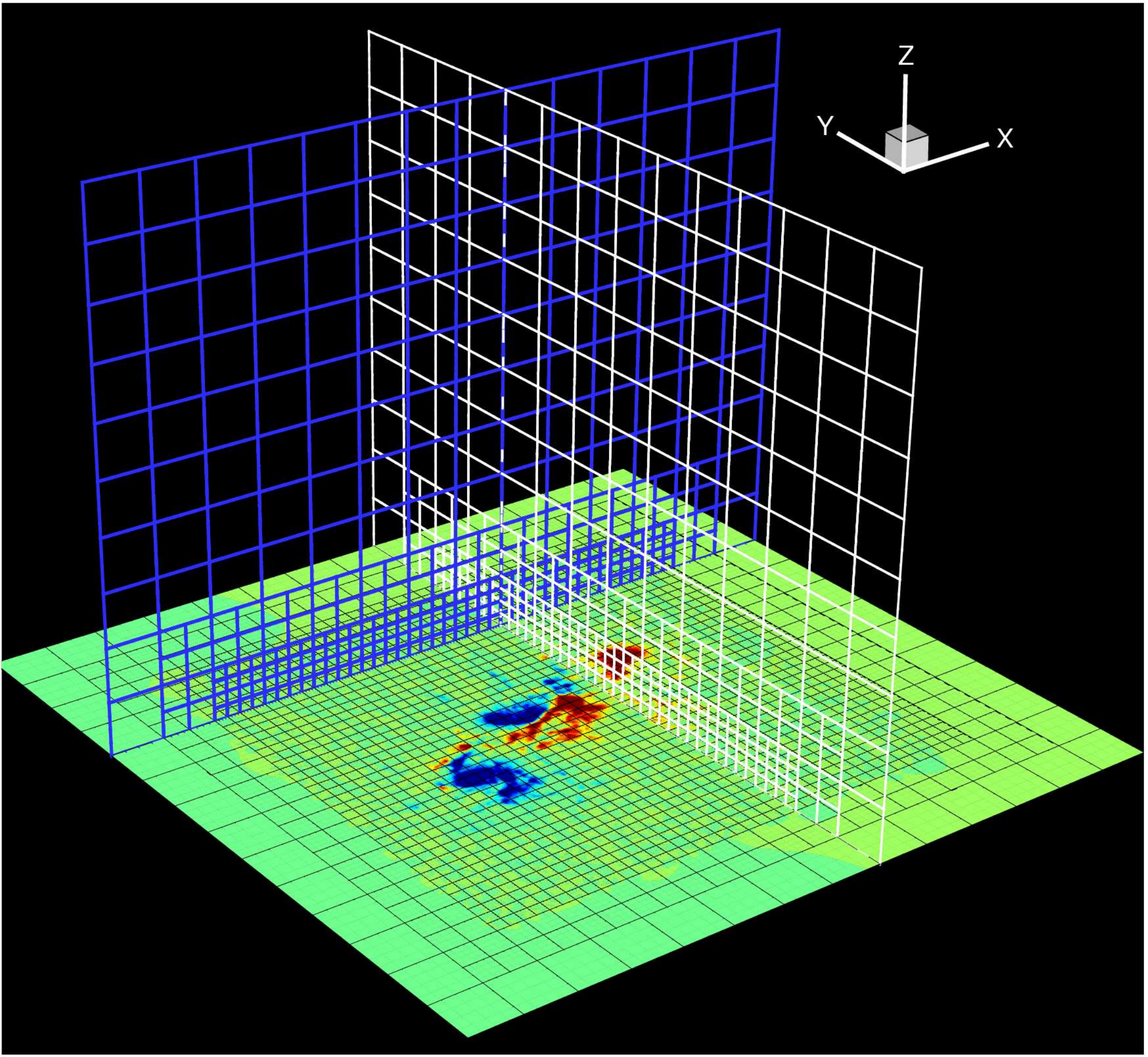}
  \caption{The grid structure: the entire volume is divided into
    blocks and each block has $8\times 8\times 8$ cells. Slices
    through the volume in three axis directions are plotted to show
    the structure of the blocks and the bottom contour map represents
    $B_{z}$ on the photosphere.}
  \label{fig:grid}
\end{figure}

Secondly, to deal with the high-resolution observation data, the
extrapolation is performed on a non-uniform grid within a block-structured,
distributed-memory parallel computational framework \citep[e.g.,][]{Jiang2012c}. Specifically, the whole computational
volume is divided into blocks with different spatial resolution, and
the blocks are evenly distributed among the processors. Within
this framework, we have lots of freedom to configure the mesh and save
computational resources comparing with a uniform grid.
As concentrated strongly in the photosphere but expanding rapidly into
the corona due to abrupt drop of the gas pressure, the coronal
field becomes smoother and weaker successively with height.
Naturally we use a grid with decreasing resolution with
height: at the bottom the grid spacing matches the resolution of the
magnetogram and up to the top of the model box, the grid spacing is
increased by, say, four times. At present we use the same resolution
in the horizontal plane, and application of adaptive resolution based
on the pattern of magnetic flux distribution is under
development. With this framework, we also add some
coarse buffer blocks around the central volume to reduce influence by
the numerical boundaries without adding much computational burden.
An example of the grid is shown in \Fig~\ref{fig:grid} .

Routinely the quantities in the computational volume is initialized by
setting $\vec B_{1}=\vec 0$ and $\vec v = \vec 0$. The potential field $\vec
B_{0}$ is obtained by a Green's function method
\citep[e.g.,][]{Metcalf2008}. The bottom boundary is incrementally fed
with the observed vector magnetogram in tens of Alfv\'en time
$\tau_{\rm A}$, while all the other numerical boundaries are fixed
with $\vec B_{1}=\vec 0$ and $\vec v=\vec 0$.

\section{Data}
\label{sec:data}


\begin{figure}[htbp]
  \centering
  \includegraphics[width=0.40\textwidth]{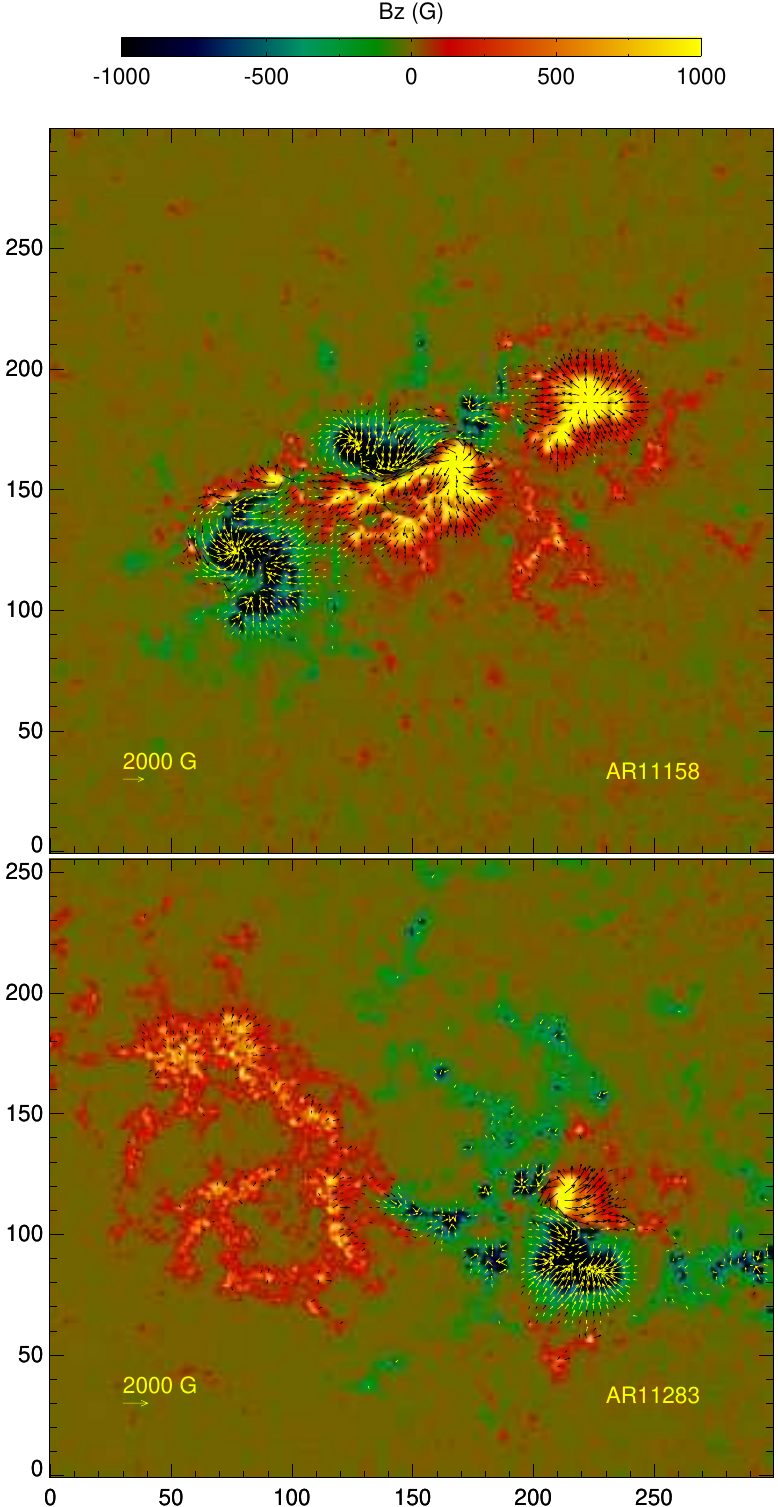}
  \caption{Vector magnetograms for AR 11158 and AR 11283. The
    background shows the vertical components with saturation values of
    $\pm 1000$~G; the vectors represent the transverse fields (above $200$~G).
	The length unit is arcsec.}
  \label{fig:vectormap}
\end{figure}

\subsection{The HMI Data}
\label{sec:hmi}

The Helioseismic and Magnetic Imager (HMI) on board the {\it Solar
  Dynamics Observatory (\SDO)} provides photospheric vector
magnetograms with a high resolution both in space and time. It
observes the full Sun with a 4K$\times$4K CCD whose spatial sampling
is 0.5 arcsec per pixel. Raw filtergrams are obtained at 6 different
wavelengths and 6 polarization states in the Fe {\sc i} 6173~{\AA}
absorption line, and are collected and converted to observable
quantities (like Dopplergrams, continuum filtergrams, and
line-of-sight and vector magnetograms) on a rapid time cadence. For
the vector magnetic data, each set of filtergrams takes 135~s to
be completed. To obtain vector magnetograms, Stokes parameters are
first derived from filtergrams observed over a 12-min interval and
then inverted through the Very Fast Inversion of the Stokes Vector
\citep{Borrero2011}. The 180$^{\circ}$ azimuthal ambiguity in the
transverse field is resolved by an improved version of the ``minimum
energy'' algorithm \citep{Leka2009}. Regions of interest with strong
magnetic field are automatically identified near real time
\citep{Turmon2010}. A detailed description on how the vector
magnetograms are produced can be found on the website
\url{//http://jsoc.stanford.edu/jsocwiki/VectorPaper}.

The magnetogram data we use here is downloaded from
\url{http://jsoc.stanford.edu/jsocwiki/VectorPaper}, where the HMI
vector magnetic field data series \texttt{hmi.B\_720s\_e15w1332} are
released for several ARs. There are two special formats,
i.e., direct cutouts and remapped images. We use the remapped
format which is more suitable for modeling in local Cartesian
coordinates, since the images are computed with a Lambert cylindrical
equal area projection centered on the tracked region. For our test, we
select two ARs, AR 11158 and AR 11283, both of which
produced X-class flares and were very non-potential. The full
resolution of the data is about 0.5$''$ per pixel and we rebin them to
1$''$ per pixel for the NLFFF modeling.

AR 11158 is a well-known target studied in many recent works for
different purposes \citep[e.g.,][]{Schrijver2011, Sun2012, Liu2012,
  Jing2012}, and was selected by \citet{Wiegelmann2012} for special
test on optimizing their extrapolation code with HMI data.
This AR has a multipolar and complex structure.
It produced several major flares and coronal mass ejections (CMEs)
during its disk passage, including the first X-class flare of cycle 24
on 2012 February 15. \Fig~\ref{fig:vectormap} (the first panel) shows
a vector magnetogram of AR 11158 taken at 20:36~UT on 2011 February
14, which will be used for our computation. This AR is well isolated
from other ones and is almost flux-balanced (see
Table~\ref{tab:quality}), with the main polarities concentrated in the
central field of view (FoV). As can be seen, the field shows a strong
shearing along the polarity inversion lines (PILs).

Another one, AR 11283, is also very eruptive, generating several X-class flares and CMEs.
We select a magnetogram taken at 22:00~UT on September 6, just
prior to a major flare at 22:20~UT. The magnetogram is shown in
the second panel of \Fig~\ref{fig:vectormap}. As an input for
extrapolation, this data is not good as AR 11158's, since the flux is
dispersed with some strong polarities almost on the edge of the
FoV. Also the total flux is not well balanced (see
Table~\ref{tab:quality}), which is unfavorable for our extrapolation.

\begin{figure*}[t!]
  \centering
  \includegraphics[width=\textwidth]{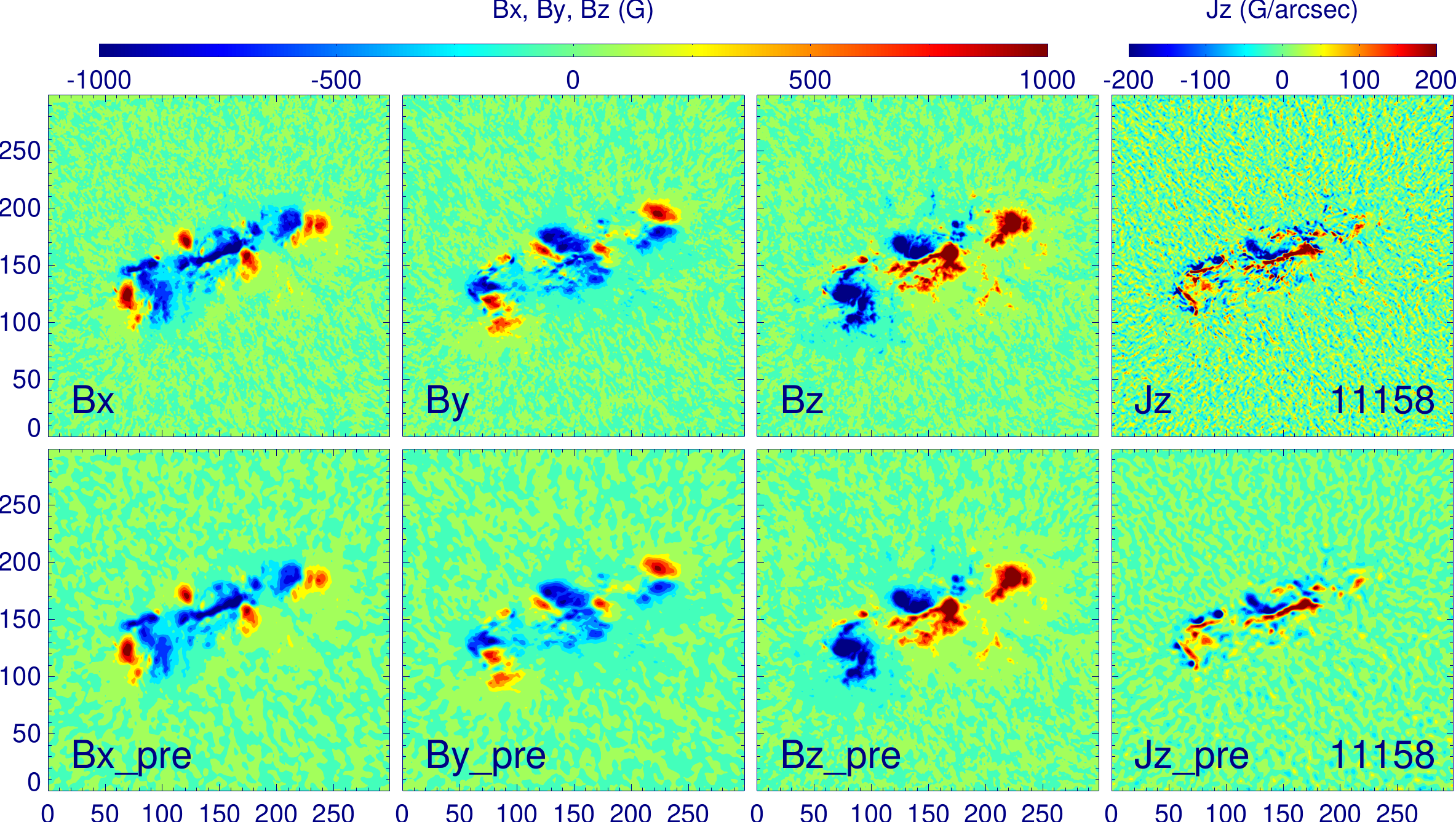}
  \caption{Raw and preprossessed magnetograms for AR 11158.}
  \label{fig:magnetogram}
\end{figure*}

\begin{figure*}[t!]
  \centering
  \includegraphics[width=\textwidth]{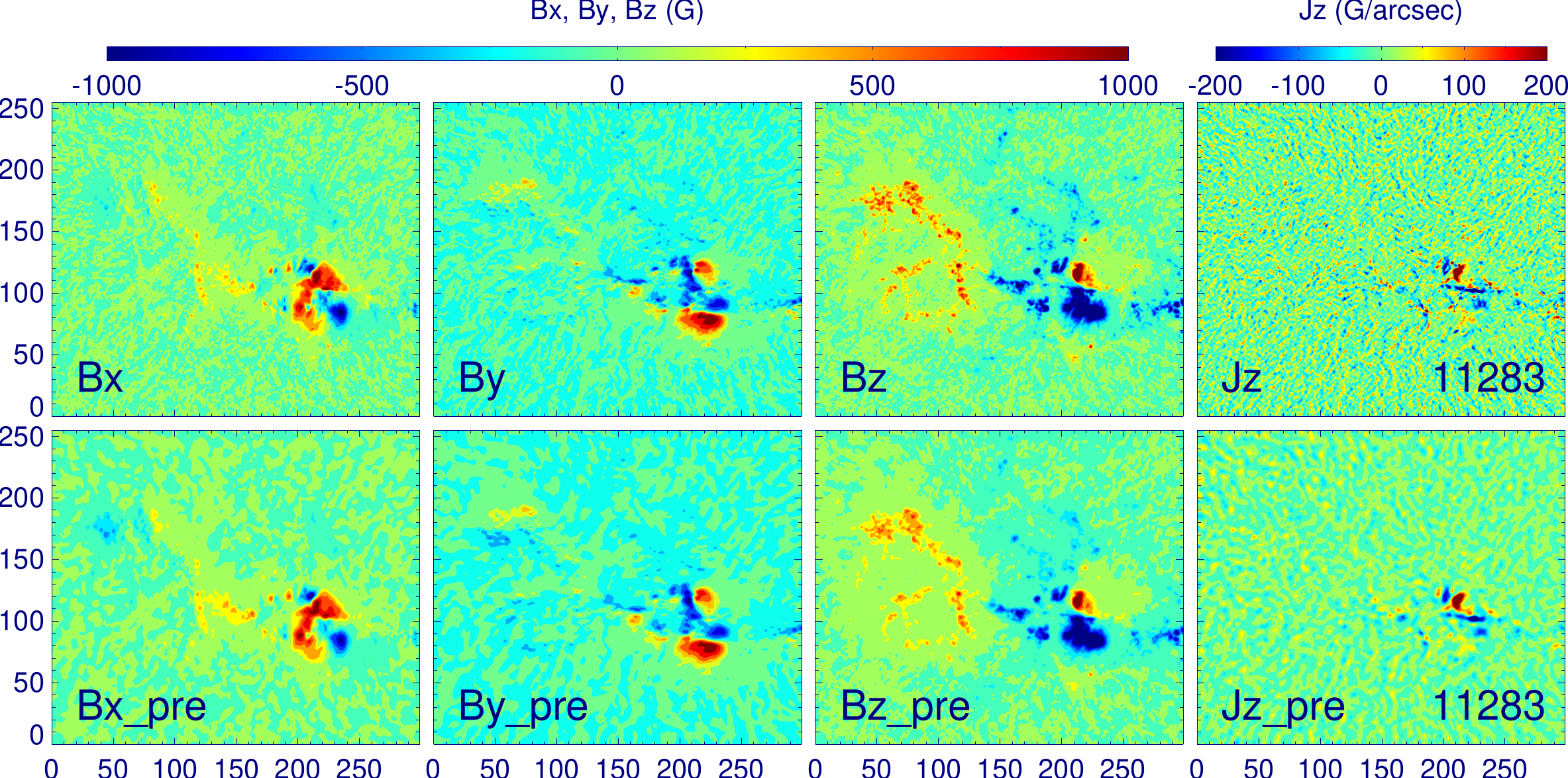}
  \caption{Raw and preprossessed magnetograms for AR 11283.}
  \label{fig:magnetogram1}
\end{figure*}

\begin{table*}[t!]
  \centering
  \begin{tabular}{lllllll}
    \hline
    \hline
    Data & $\epsilon_{\rm flux}$ & $\epsilon_{\rm force}$ &
    $\epsilon_{\rm torque}$ & $S_{x}$ & $S_{y}$ & $S_{z}$\\
    \hline
    AR 11158\\
    Raw map           & 0.017 &   0.072  &    0.073 & 5.05E-03  &6.12E-03  &1.05E-03\\
    Preprocessed map  & 0.019 &   0.002  &    0.002 & 7.83E-05  &1.01E-04  &8.74E-05\\
    Numerical potential & 0.019 & 0.001  &    0.002 & 8.78E-05  &8.42E-05  &8.74E-05\\
    \hline
    AR 11283\\
    Raw map          & -0.111 &   0.193  &    0.143 & 1.12E-02 & 1.15E-02 & 1.94E-03 \\
    Preprocessed map & -0.119 &   0.012  &    0.017 & 1.73E-04 & 1.95E-04 & 1.79E-04 \\
    Numerical potential  & -0.119 & 0.011&    0.017 & 1.98E-04 & 1.63E-04 & 1.79E-04\\
    \hline
 \end{tabular}
 \caption{Quality of the magnetograms. }
 \label{tab:quality}
\end{table*}

\subsection{Data Preprocessing}
\label{sec:pre}

Generally, the raw magnetogram cannot be inputted directly into the
NLFFF code because the intrinsic non-force-freeness of the
photospheric field violates the force-free assumption. According to
the derivation by \citet{Molodenskii1969} and \citet{Aly1989}, an
ideally force-free magnetogram should fulfill the following conditions:
\begin{eqnarray}
  \label{eq:m1}
  \int_{S} B_{z} \rmd x\rmd y = 0,\ \
  F_{x} = \int_{S}B_{x}B_{z}\rmd x\rmd y = 0,\nonumber\\
  F_{y} = \int_{S}B_{y}B_{z}\rmd x\rmd y = 0,\ \
  F_{z} = \int_{S}E_{B}\rmd x\rmd y = 0,\nonumber\\
  T_{x} = \int_{S}yE_{B}\rmd x\rmd y = 0,\ \
  T_{y} = \int_{S}xE_{B}\rmd x\rmd y = 0, \nonumber\\
  T_{z} = \int_{S}(yB_{x}B_{z}-xB_{y}B_{z}) \rmd x\rmd y = 0
\end{eqnarray}
where $E_{B} = B_{x}^{2}+B_{y}^{2}-B_{z}^{2}$. These expressions are
derived from the volume integrals of the total divergence, magnetic
force and torque of an ideally force-free field
\begin{equation}
  \label{eq:m11}
  \int_{V} \divB \rmd V = 0, \ \
  \int_{V} \vec j\times \vec B \rmd V = \vec 0,\ \
  \int_{V} \vec r\times (\vec j\times \vec B) \rmd V = \vec 0
\end{equation}
(by using Gauss' divergence theorem, the volume integrals~\ref{eq:m11} can be transformed
to the surface integrals~\ref{eq:m1}). In the case of the coronal field, the surface
integrals of \Eq~(\ref{eq:m1}) are usually restricted within the bottom
magnetogram since the contribution from other boundaries can be
neglected. To assess the real data with respect to the force-free condition,
three parameters are usually computed as \citep{Wiegelmann2006b}
\begin{eqnarray}
  \label{eq:ffq}
  \epsilon_{\rm flux} = \frac{\int_{S} B_{z} \rmd x\rmd y}{\int_{S} |B_{z}|
    \rmd x\rmd y},\ \
  \epsilon_{\rm force} = \frac{|F_{x}|+|F_{y}|+|F_{z}|}{\int_{S} P_{B}
    \rmd x\rmd y},\nonumber\\
  \epsilon_{\rm torque} = \frac{|T_{x}+|T_{y}|+|T_{z}|}
  {\int_{S} \sqrt{x^{2}+y^{2}}P_{B} \rmd x\rmd y}
\end{eqnarray}
where $P_{B} = B_{x}^{2}+B_{y}^{2}+B_{z}^{2}$. Small values of these
quantities, e.g., $\epsilon_{\rm flux}, \epsilon_{\rm force},
\epsilon_{\rm torque} \ll 1$ indicate a good input for the NLFFF
modeling. Table~\ref{tab:quality} shows that for AR 11158, the
force-free condition is satisfied quiet well with flux almost balanced and
$\epsilon_{\rm force}, \epsilon_{\rm torque}$ less
than $0.1$; while for AR 11283, it is worse. Note that the flux
non-balance will pose a negative effect on the extrapolation (see
Section~\ref{sec:resu}).

Besides the non-force-freeness, the observed data contains measurement
noise which is also unfavorable for practical implementation of
extrapolation. To this end, preprocessing of the raw magnetogram has
been proposed by \citet{Wiegelmann2006b} to remove the force and noise
for providing better input for NLFFF modeling. To be consistent with
our extrapolation code using a magnetic splitting form, we
recently developed a new code for magnetogram preprocessing
\citep{Jiang2012SoPsubmit}, in which
the vector magnetogram is also split into a potential field part and a
non-potential field part and we deal with the two parts
separately. Preprocessing of the potential part is simply performed by
taking the data sliced at a plane about $400$~km above the
photosphere\footnote{We choose such height because the field becomes
  force-free in the chromosphere roughly 400~km above the photosphere
  according to \citep{Metcalf1995}. Our preprocessing code is designed
  to modify the photospheric magnetogram to mimic a force-free
  chromospheric magnetogram at such height.} from the 3D
potential field which is extrapolated from the
observed vertical field. Then the non-potential part is modified and
smoothed by an optimization method to fulfill the constraints of total
magnetic force-freeness and torque-freeness, which is similar to the
method proposed by \citet{Wiegelmann2006b}. We have paid particular attention to
the extents the force is needed to be removed and
the smoothing can be performed. As for practical computation based on
numerical discretization, an accurate satisfaction of force-free
constraints is apparently not necessary. Also the extent of the
smoothing for the data needs to be carefully determined, if we want to
mimic the expansion of the magnetic field from the photosphere to some
specific height above. We use the values of force-freeness and
smoothness calculated from the preprocessed potential-field part as a
reference to guide the preprocessing of the non-potential field part,
i.e., we require that the target magnetogram has the same level of
force-freeness and smoothness as its potential part. These
requirements can restrict well the free parameters, i.e., the weighted
factors in the optimization function.

The results of preprocessed data are also given in
Table~\ref{tab:quality} together with results of their numerical
potential-field part. For both magnetograms, the preprocessing reduces the
parameters $\epsilon_{\rm force}$ and $\epsilon_{\rm torque}$ by more
than one order of magnitude, making the residual force around the
level of numerical error (i.e., the parameters are close to those of the
numerical potential field). The parameters $S_{x},S_{y},S_{z}$ in
the table measure the smoothness of the components $B_{m}$
($m=x,y,z$), which are defined by \citep[see][]{Jiang2012SoPsubmit}
\begin{equation}
  S_{m} = \sum_{\rm p}\left[(\Delta B_{m})^{2}\right]/
  \sum_{\rm p}\left[(\overline{\Delta} B_{m})^{2}\right]
\end{equation}
where the summation $\sum_{\rm p}$ is over all the pixels of the
magnetogram, and $\Delta$ is a usual five-point 2D-Laplace operator, i.e.,
for pixel $(i,j)$
\begin{eqnarray}
  \Delta B_{i,j} \equiv B_{i+1,j}+B_{i-1,j}+B_{i,j+1}+B_{i,j-1}-4B_{i,j},\nonumber \\
  \overline{\Delta}B_{i,j} \equiv B_{i+1,j}+B_{i-1,j}+B_{i,j+1}+B_{i,j-1}+4B_{i,j}.
\end{eqnarray}
As shown the smoothness of the preprocessed data is very close to those
of their potential part, and is consistent among three components,
which is unlike the raw data with very different smoothness for different components.
Smoothing of the data can be clearly
seen by comparing the raw and preprocessed magnetograms as shown in
\Fig~\ref{fig:magnetogram}, and especially in the $J_{z}$ map which
shows the random noise is suppressed effectively (but not totally).

Readers should be reminded that the constraints of \Eq~(\ref{eq:m1})
are only necessary conditions, but not sufficient for an ideally
force-free magnetogram, meaning that the magnetogram may still
contains force even these conditions are satisfied. What we can say is
that the preprocessed magnetogram is more suitable for NLFFF modeling
than the raw data, but not a completely consistent boundary condition.

\section{Results}
\label{sec:resu}

\begin{figure*}[htbp]
  \centering
  \includegraphics[width=\textwidth]{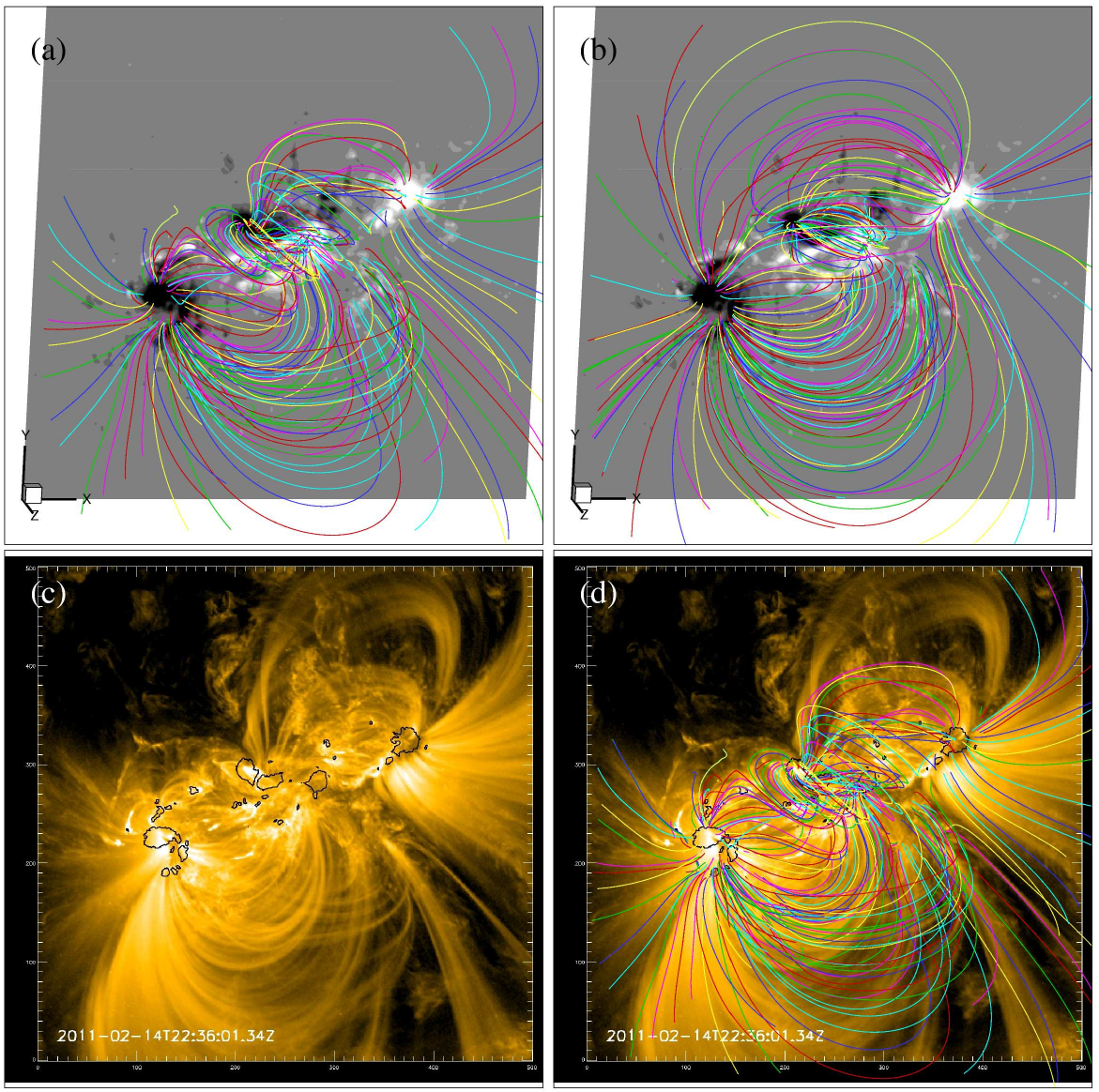}
  \caption{Comparison of extrapolation field lines with AIA
    $171$~{\AA} loops for AR 11158: the NLFFF lines (a), the potential
    field lines (b), the AIA image (c) and NLFFF lines overlaying the
    AIA image (d). Contour lines for $\pm 1000$~G (the black curves)
    of LoS photospheric field are over-plotted on the AIA images, and
    for all the panels the field lines are traced from the same set of
    footpoints on the bottom surface.}
  \label{fig:AR11158_AIA}
\end{figure*}

\begin{figure*}[htbp]
  \centering
  \includegraphics[width=\textwidth]{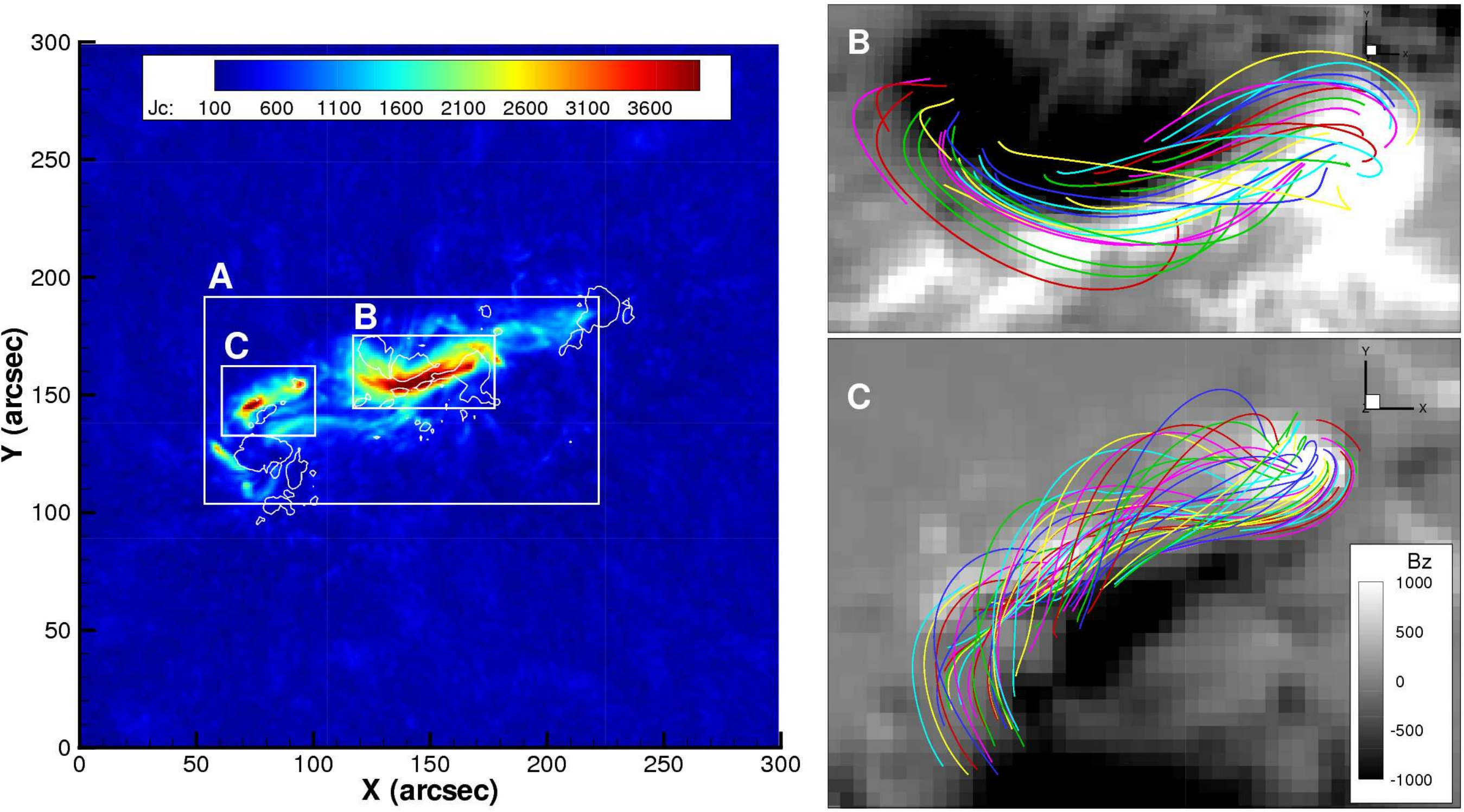}
  \caption{Strong-current regions and their magnetic structures for AR
    11158. Left panel is image of the vertical integral of current
    density, i.e., $J_{\rm C} = \int |\vec J| dz$ where the current is
    calculated by $\vec J = \crlB$ with a unit of G~pixel$^{-1}$; Regions
    of strong current are outlines by the boxes A, B and C. Right
    panels show the magnetic structures of the strong-current regions
    of B and C.}
  \label{fig:AR11158_Jc}
\end{figure*}

\begin{table*}[htbp]
  \centering
  \begin{tabular}{llllrrrrr}
    \hline
    \hline
    Region & CWsin & C$^{2}$Wsin & $\langle |f_{i}|\rangle$ & $E_{\rm tot}$ &
    $E_{\rm pot}$ & $E_{\rm free}$ & $E_{\rm free}/E_{\rm pot}$ & $E_{\rm free}/(E_{\rm free})_{\rm full}$ \\
    \hline
    Raw  \\
    Full &  0.32  &  0.18  &  6.59E-04  &  10.9  &  8.97  &  1.93  & 22\% & 100\% \\
    A    &  0.18  &  0.11  &  1.25E-03  &  8.3   &  6.24  &  2.06  & 33\%& 107\% \\
    B    &  0.07  &  0.06  &  7.17E-04  &  3.24  &  2.09  &  1.15  & 55\%& 60\%  \\
    C    &  0.16  &  0.11  &  1.64E-03  &  0.67  &  0.47  &  0.20  & 43\%& 10\%  \\
    \hline
    Preprocessed \\
    Full &  0.30  &  0.14  &  6.18E-04  &  9.83  &  8.18  &  1.65  & 20\%&100\% \\
    A    &  0.17  &  0.10  &  1.20E-03  &  7.54  &  5.61  &  1.93  & 34\%&117\% \\
    B    &  0.06  &  0.05  &  7.20E-04  &  2.97  &  1.81  &  1.16  & 64\%&70\%  \\
    C    &  0.16  &  0.12  &  1.67E-03  &  0.59  &  0.42  &  0.17  & 40\%&10\%  \\
    \hline
  \end{tabular}
  \caption{Results of the metrics for AR 11158.
    Full region is extrapolation box of $[0,300](x)\times [0,300](y)\times [0,150](z)$.
    Region A is $[53,222]\times [104,191]\times [0,50]$;
    Region B is $[116,181]\times  [145,175]\times [0,30]$;
    Region C is $[62,102]\times [132,161]\times [0,20]$. The FoVs of
    the regions are shown by the boxes in \Fig~\ref{fig:AR11158_Jc}.
    The energy unit is $10^{32}$~erg.}
  \label{tab:AR11158}
\end{table*}

\subsection{AR 11158}
\label{sec:AR11158}

The observed coronal loops in X-ray and EUV images give us a proxy of
the magnetic field line geometry and are thus a good constraint for
the magnetic field model besides the vector magnetograms
\citep[e.g.,][]{Aschwanden2012ApJ,Malanushenko2012}. In
\Fig~\ref{fig:AR11158_AIA} we compare the extrapolated field with
coronal loops observed by {\it \SDO}/AIA in the wavelength of 171
{\AA}, in which the loops are visible the best than in other
channels. The field lines are traced from the photosphere at locations
selected roughly according to the footpoints of the visible bright
loops, and are rendered with different colors for a better looking.  The
view angle of the field lines is aligned with the AIA image. We plot
the field lines and the AIA image both alone and overlaid for a better
inspecting of the result. As shown from an overview of the images,
most of the extrapolated field lines closely resemble the observed
loops. At the central region the field lines are strongly sheared
along the major PIL and slightly twisted (as compared with the
potential solution), indicating the existence of strong electric
currents along the field lines. In the left panel of
\Fig~\ref{fig:AR11158_Jc}, we plot an image of vertical integral of the
current density in the extrapolation volume, and the strong-current
regions are outlined by the boxes (labeled as A, B and C). Note that
the currents are strongly localized within the central region B and a
smaller region C. The magnetic structures of the strong-current
regions are shown in the right panel of \Fig~\ref{fig:AR11158_Jc}. As
can be seen, the twist of the field lines in region C is much stronger
than that in region B. The results support observation studies which
show that a filament related with a X-class flare and CME exists in
the core region B \citep{Sun2012}, and there are small eruptions in
region C due to flux emergence \citep{Sun2012b}.

Nevertheless, we note that the misalignments between the modeling
field lines and the observation are also obvious, especially, the
large loops near the north-west boundary of the FoV. There are several
reasons for the misalignments: first, a local Cartesian coordinate
system is not adequate to include the large loops which obviously
need spherical geometry; second, the FoV of magnetogram may be not
large enough to properly characterize the entire relevant current
system, which again needs the curvature of the Sun's surface to be
taken into account; third, it is not easy to locate precisely the
photospheric footpoints of different loops that spread apart
distinctly in the corona but are rooted very closely in the
photosphere; fourth, the coronal field may be rather dynamic, e.g.,
expands or oscillates due to eruptions, which makes the static
extrapolation fail. We know that this
visual comparison between the model result and observation is very
preliminary, and more critical comparison is requred, for example,
with the 3D loops reconstructed with multi-points observation
\citep{DeRosa2009,Aschwanden2011}.

Routinely, we check the quality of the numerical result by computing
several metrics. The force-freeness of the extrapolation data is
usually measured by a current-weighted sine metric (CWsin) defined by
\begin{equation}
  {\rm CWsin} \equiv
  \frac{\int_{V}{J}\sigma dV}{\int_{V}{J} dV};
  \ \
  \sigma =  \frac{{\vec J}\times{\vec B}}{{J}{B}},
\end{equation}
where $B=|\vec B|$, $J=|\vec J|$ and $V$ is the computational
volume. We also compute a current-square-weighted sine metric
(C$^{2}$Wsin) similarly defined by
\begin{equation}
  {\rm C^{2}Wsin} \equiv
  \frac{\int_{V}J^{2}\sigma dV}{\int_{V}J^{2} dV},
\end{equation}
with more weight on the strong-current regions. The
divergence-freeness is measured by $\langle |f_{i}|\rangle$
\begin{equation}
  \langle |f_{i}|\rangle =
  \frac{1}{V}\int_{V}\frac{\divB}{6B/\Delta x} dV.
\end{equation}
We care about different energy contents, i.e., the total energy
$E_{\rm tot}$, the potential energy $E_{\rm pot}$ and the free energy
$E_{\rm free}$
\begin{equation}
  E_{\rm tot} = \int_{V} \frac{B^{2}}{8\pi} dV,\ \
  E_{\rm pot} = \int_{V} \frac{B_{\rm pot}^{2}}{8\pi} dV,\ \
  E_{\rm free} = E_{\rm tot}-E_{\rm pot}
\end{equation}
where $B_{\rm pot}$ is the potential field strength. Results of the
metrics are given in Table~\ref{tab:AR11158} for extrapolations from
both the raw and preprocessed magnetograms. We compute the metrics for
four different regions including the full extrapolation box and the
subregions A, B, and C as outlined in
\Fig~\ref{fig:AR11158_Jc}.

For the full region, our results of the current-weighted sine is $\sim
0.3$, which means the mean misalignment angle between the magnetic
field and current is about $17^{\circ}$. Such value is much larger
than those from our previous benchmark tests using ideal or synthetic magnetograms
\citep[which are $\sim 0.1$~(6$^{\circ}$) or smaller, see][]{Jiang2012apj},
but is comparable to previously reported results by other NLFFF
codes on real magnetograms (e.g., the average CWsin by various NLFFF
codes applied to AR 10930 \citep{Schrijver2008} and AR 10953
\citep{DeRosa2009} are 0.36 and 0.28, respectively). With such a large
misalignment angle, the result seems to be far away from an exactly
force-free solution which has a zero misalignment angle. However it
should be noted that the metric CWsin may not be a good monitor for
numerical solutions, which unavoidably have random numerical errors
because of limited resolution. As a simple example, CWsin
is close to $1$ even for a potential field solution computed by
Green's function method or other numerical realization. The reason is
that the numerical difference, used for computing the current $\vec J
= \curl \vec B$ from $\vec B$, gives small but finite currents, whose
directions are randomly from 0$^{\circ}$ to 180$^{\circ}$, thus an
average of the full volume should give a misalignment angle of $\sim
90^{\circ}$. The noise in the observation data, mainly in the weak
field regions, is also a major source for the random numerical
errors. In these regions, the actual magnetic elements are probably
smaller than the observed or numerical pixel size, and the field
directions generally exhibit a random pattern on the image. To reduce
such errors in computing the metric, we can either put larger weight of current
(e.g., use C$^{2}$Wsin) or compute CWsin within the strong-current
subregions only. As is expected, the misalignment angle decreases
significantly by measuring in this way. For the full region,
C$^{2}$Wsin are only half of CWsin. For the subregions,
CWsin are also only half or less for the full region,
reaching the level of those from the benchmark tests
\citep{Jiang2012apj}. In particular, the misalignment angle is only
about $4^{\circ}$ in subregion B, showing that the force-free
assumption is modeled very well.

Regarding the energy contents, it is interesting to note that the free
energy of subregion A exceeds that of the full region, meaning that
the free energy content in the full volume excluding subregion A is
negative. This is, however, not surprising as we know that any
sub-volume energy content of the non-potential field may be lower than
the potential energy \citep[e.g.,][]{Mackay2011,Jiang2012c}.
Also the measurement error may result in this negative free energy since it is very small compared to the total free energy.
No matter which case is true, it can be
clearly seen that the spatial distribution of free energy is largely
co-spatial with that of the current, since the subregion A contains
most of the currents of the whole volume. This confirms that the free
energy in the corona is actually stored by the current-carrying field
(where non-potentiality is strong), but not necessarily in the
magnetic flux concentrations.

Finally, we compare the results extrapolated from the raw and
preprocessed magnetograms. Inspecting of the force-freeness and
divergence-freeness metrics shows that the improvement by preprocessing
is negligible. This is because the
raw data already satisfies the boundary force-free conditions
well. Due to the smoothing, the result for the preprocessed data
gives slightly lower energy contents than those for the raw data.

\begin{figure*}[htbp]
  \centering
  \includegraphics[width=\textwidth]{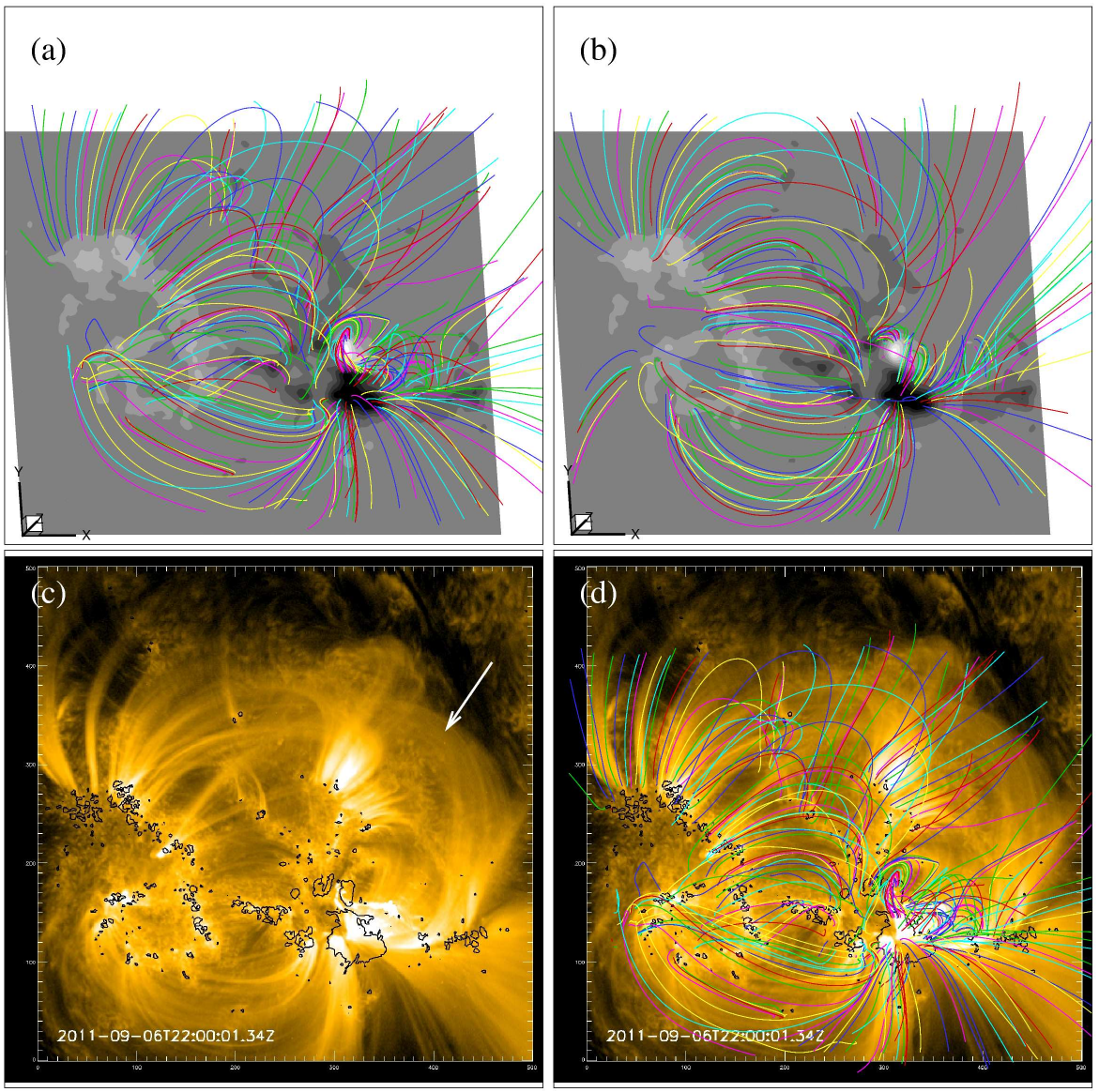}
  \caption{Same as \Fig~\ref{fig:AR11158_AIA} but for AR 11283.}
  \label{fig:AR11283_AIA}
\end{figure*}

\begin{figure*}[htbp]
  \centering
  \includegraphics[width=0.8\textwidth]{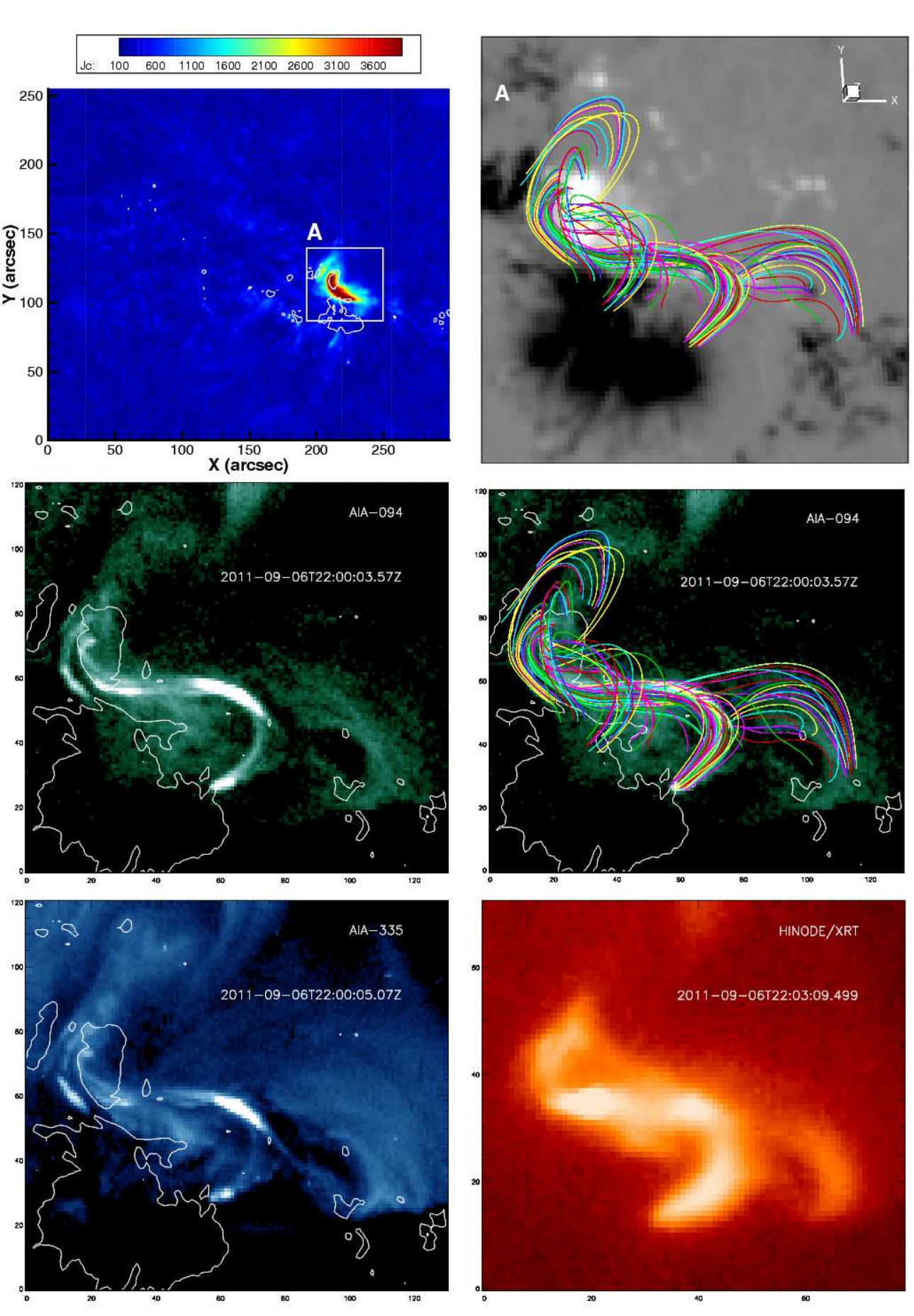}
  \caption{Strong-current region and its magnetic structure for AR
    11283. The upper-left panel is image of the vertical integral of
    current density, i.e., $J_{\rm C} = \int |\vec J| dz$; and
    upper-right panel shows the magnetic structure of subregion
    A, where a sigmoid, shown in the following panels,
    is observed clearly by \SDO/AIA and Hinode/XRT.}
  \label{fig:AR11283_Jc}
\end{figure*}

\begin{table*}[htbp]
  \centering
  \begin{tabular}{llllrrrrr}
    \hline
    \hline
    Region & CWsin & C$^{2}$Wsin & $\langle |f_{i}|\rangle$ & $E_{\rm tot}$ &
    $E_{\rm pot}$ & $E_{\rm free}$ &  $E_{\rm free}/E_{\rm pot}$ & $E_{\rm free}/(E_{\rm free})_{\rm full}$ \\
    \hline
    Raw  \\
    Full &  0.40  &  0.24  &  9.65E-04  &  5.94  &  5.58  &  0.46 &8\% & 100\% \\
    A    &  0.15  &  0.09  &  3.69E-03  &  1.86  &  1.33  &  0.53 &40\% & 115\% \\
    \hline
    Preprocessed \\
    Full &  0.32  &  0.18  &  8.28E-04  &  6.10  &  5.12  &  0.98 &19\% & 100\% \\
    A    &  0.13  &  0.09  &  1.76E-03  &  2.05  &  1.18  &  0.87  &74\%& 89\%  \\
    \hline
  \end{tabular}
  \caption{Results of the metrics for AR 11283.
    Full region is extrapolation box of $[0,300](x)\times [0,256](y)\times [0,150](z)$.
    Region A is $[193, 251]\times [86,140]\times [0,30]$. The FoVs of
    the regions are shown by the box in \Fig~\ref{fig:AR11283_Jc}.
    The energy unit is $10^{32}$~erg.}
  \label{tab:AR11283}
\end{table*}

\subsection{AR 11283}
\label{sec:AR11283}

\Fig~\ref{fig:AR11283_AIA} compares the AIA 171~{\AA} loops with the
reconstructed field in the same way as \Fig~5. For this
AR, the NLFFF model appears to perform only slightly
better than the potential model (there are some loops even worse produced by the NLFFF model than
the potential model near the north-east boundary of the FoV). The
clearest misalignment with the observation is the large closed loop
pointed by the arrow in AIA image. This group of loops are failed to be
recovered by both the potential and force-free models which give open
field lines instead. This, however, is not unexpected since the
inputted magnetogram has flux unbalanced by $-10\%$. So there must
be field lines from the negative polarity opening in the FoV. The
reason for this flux unbalance may be that the positive flux in the
east is rather dispersed (much more than the negative polarity), and
thus properly be underestimated by the observation.

Near the major polarity the structure of the loops is very
complex and the extrapolated field shows highly sheared and twisted
structures, indicating a significant non-potentiality there. Actually
this was the site of the flare and filament eruption. The
distribution of the vertically integrated current shows a strong
concentration of current in this region, as denoted by A in
\Fig~\ref{fig:AR11283_Jc}. In the same figure, we show the local field
structure and the observations from different wavelengths of much
higher temperature than 171~{\AA}. The magnetic field exhibits a
multi-flux rope configuration. The most remarkable structure is a
sigmoid, i.e., the S-shaped loop in the AIA 94~{\AA} and 335~{\AA}
images. The sigmoid can be seen most clearly in the AIA 94~{\AA}
wavelength ($6.3$~MK) with rather thin but enhanced shape, and is
also well shaped in the soft X-ray image taken by {\it Hinode}/XRT. In
the fourth panel of the figure the field lines are plotted overlying
on the AIA 94~{\AA} image. It demonstrates that our extrapolation has
recovered the sigmoid rather precisely, at least in morphology (see
the precise alignment of the field lines with the shape of the
sigmoid). The distribution of the current also resembles roughly the
shape of the sigmoid, suggesting that the enhancement of EUV and X-ray
emission associated with the sigmoid is made by the strong
field-aligned current via Joule heating of the plasma. This sigmoid
locates between the major positive and negative polarities and the
currents reside mostly in the north-east part, as shown by the current
distribution. The twist of the sigmoid field lines is
not strong as modeled in other cases such as \citet{Roussev2012} or
\citet{Savcheva2012}, and this sigmoid is composed of a single flux
rope, which is also different from their results with two flux ropes
or double-J shaped current pattern. The observation and modeling
suggest that there seems to be another flux rope overlying the
sigmoid, and the flare and CME may be resulted by the eruptions of these flux
ropes, which is left for future study.

Similarly, we compute the metrics of force-freeness and
divergence-freeness for both the full region and the subregion and the
results are given in Table~\ref{tab:AR11283}. By comparing the results
using the raw and preprocessed data, we find that evidently
the preprocessed result is closer to force free, especially of the
full region for which the raw data gives CWsin$\sim 0.4$
($24^{\circ}$) while the preprocessed data gives CWsin$\sim 0.3$
($17^{\circ}$). Thus for this AR the preprocessing indeed improves
the extrapolation greatly. Also the
divergence is reduced by the preprocessing.
It is noticeable that the total energy content
is doubled by the preprocessing, reaching $10^{32}$~erg. But even this improvement
of the free energy is likely to underestimate the actual value, considering that a
X-class flare and CME erupted immediately \citep{FengL2013}. Still the
current is strongly localized and the free energy is concentrated
within the strong-current region, i.e., subregion A, which occupies
only less than one percent of the full volume, but contains most of
the free energy.

\begin{figure*}[htbp]
  \centering
  \includegraphics[width=0.8\textwidth]{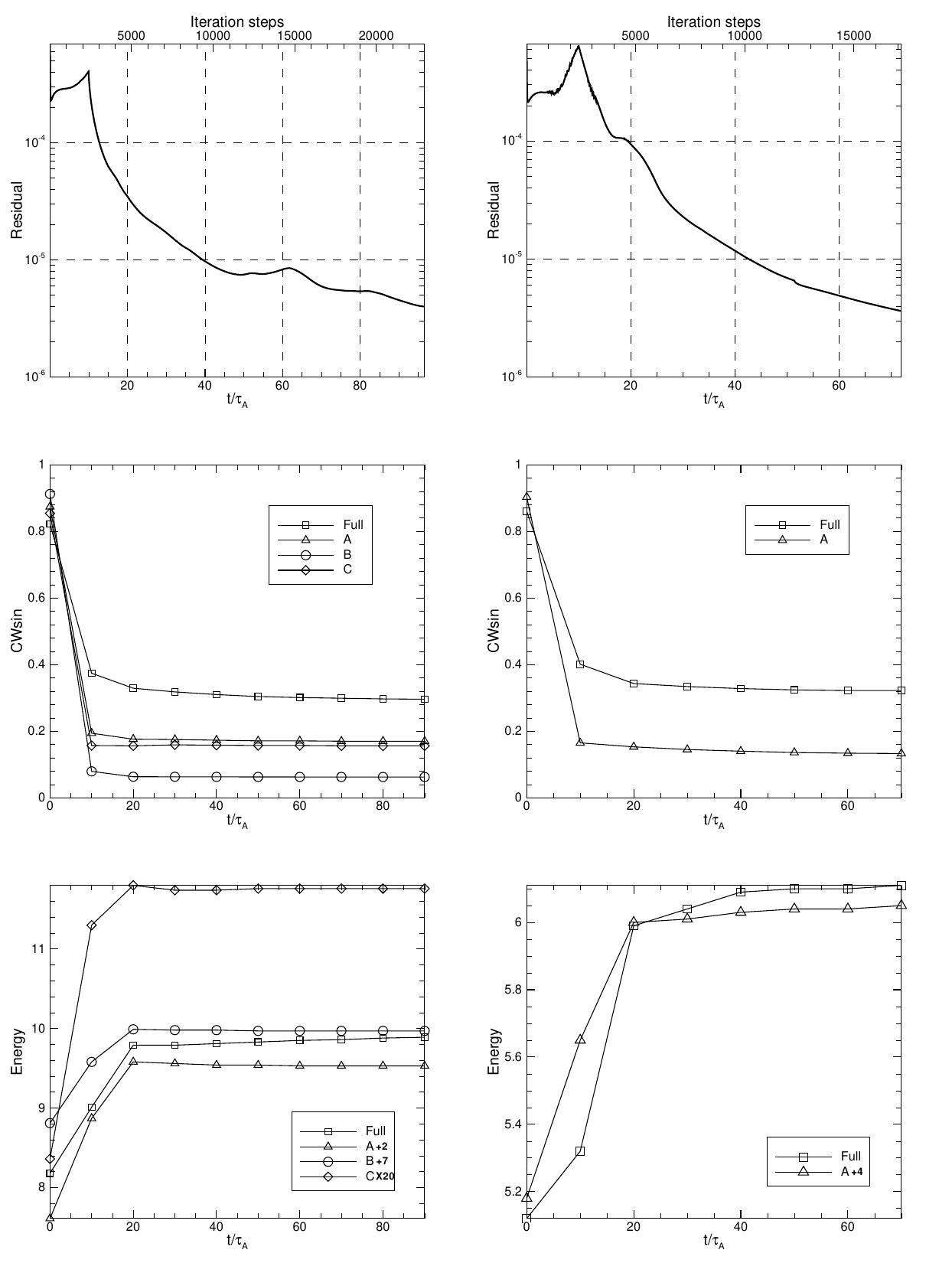}
  \caption{Convergence of computations: temporal evolution of the
    residual, the metric CWsin and the total energy content in the
    iteration process for AR 11158 (left column) and AR 11283 (right
    column). Note that for a better plot for the energy contents of
    different regions, their values are scaled properly as shown in
    the line legends.}
  \label{fig:Conv}
\end{figure*}

\subsection{Convergence Study}
\label{sec:conv}

It is important to monitor the relaxation process to study whether the
iteration converges, since there is no theory to guarantee
this. Here we study the convergence process of the computations by
temporal evolution of several monitors, including the residual of
field between two successive iterations
\begin{equation}
  \label{eq:res}
  {\rm res}^{n}(\vec B) = \sqrt{\frac{1}{3}\sum_{\delta = x,y,z}
    \frac{\sum_{i}(B_{i\delta}^{n}-B_{i\delta}^{n-1})^{2}}{\sum_{i}(B_{i\delta}^
      {n})^{2}}}
\end{equation}
(where $n$ denotes the iteration step), the metric CWsin and the total
energy content. We record the residual by every ten steps and compute
CWsin and total energy by every ten $\tau_{\rm A}$. Results for
extrapolation of both ARs are plotted in \Fig~\ref{fig:Conv}. As can
be seen, the system converges smoothly and fast. During
the first 10~$\tau_{\rm A}$, the residual keeps increasing because the
transverse field is inputted at the bottom continuously, which drives
the system away from the initial potential field. After this driving
process, the
residual drops immediately, indicating a fast relaxation of the
system. With about 40 $\tau_{\rm A}$ (nearly 10000 iterations),
the residual is already reduced to $\sim 10^{-5}$, and all the metrics and
energy almost stagnate afterward. Thus the
computations can actually be terminated once the residual is below
$10^{-5}$, which is consistent with our previous studies for benchmark
cases \citep{Jiang2012apj,Jiang2012apj1}. It is also noteworthy that
the convergence process is rather smooth, without any obvious
oscillation or abrupt variation of the residual or the metrics, so
the iteration is ``safe''. This is a good feature of our code over
other iteration codes for extrapolation, e.g., the
\citet{Valori2007}'s magnetofrictional code or the
\citet{Wheatland2006}'s Grad-Rubin-like code, which usually show strong
oscillatory in the iteration or even fail
to converge occasionally \citep{Schrijver2008,DeRosa2009}.

\section{Conclusions}
\label{sec:colu}

In this paper we have applied the CESE--MHD--NLFFF code to the
\SDO/HMI vector magnetograms. Two ARs are sampled for the test,
AR~11158 and AR~11283, both of which produced X-class
flares and were very non-potential. We compared the results with the
\SDO/AIA images, showing that the reconstructed field lines resemble
well most of the plasma loops, which is a basic requirement for an
applicable NLFFF modeling code \citep{DeRosa2009}. Because the
magnetic flux of the AR~11283 magnetogram is not well balanced, the
extrapolation of the large scale field appears not as good as that for
AR~11158. Observation shows that in the core regions of the ARs there
were filament or sigmoid which are important precursors of
eruptions like flares and CMEs. We also found in these places, there
were highly-sheared and twisted field lines, i.e., flux ropes, which
contain strong field-aligned currents and plenty of
non-potential energy, and our extrapolations recovered indeed well
those observed features, especially the sigmoid in AR~11283. By
computing the metric CWsin which measures mean value of
misalignment between the magnetic field and electric current, we
found that, the force-free constraint is fulfilled very well in the
strong-field regions (CWsin $\approx 0.1$, misalignment about
$6^{\circ}$) but apparently not that well in the weak-field regions
(CWsin $\approx 0.3$, misalignment about $17^{\circ}$) because of the
data noise and the numerical errors of the small currents. The energy
contents of our results are also consistent with the previous
computations \citep[with respect to the AR~11158,
e.g.,][]{Wiegelmann2012,Sun2012}. In summary our extrapolation code
can be used as a viable tool to study the 3D magnetic field in the
corona.

We developed the CESE--MHD--NLFFF code not only for field extrapolation,
but also as a sub-program for the project of data-driven MHD modeling
of the ARs, the eruptions and their dynamic evolutions in the global
corona using continuously-observed data on the photosphere. At present
numerical MHD investigations of the solar eruptions
\citep{Amari2003, MacNeice2004, Aulanier2009, Fan2010, Torok2011,
  Roussev2012} are mostly based on idealized magnetic configurations
without constrained by real observations. A step forward of
understanding what really happens in the solar eruptions, certainly
necessitates the observation-constrained numerical model. For example,
considering that NLFFF extrapolation can recover highly-sheared
magnetic arches and twisted flux ropes, which are basic building
blocks of many eruption models
\citep[e.g.,][]{Torok2005,Aulanier2009}, utilizing the extrapolated
field from real magnetograms can obviously provide much more
realistic initial inputs than those idealized models like
\citet{Titov1999}'s flux rope model. Our future work is to
input the extrapolation field as an initial condition into the
data-driven full MHD model \citep{Jiang2012c,Feng2012apj}, along with
the surface plasma flows derived from time-series of photosphere
magnetograms \citep[e.g.,][]{Liu2012} as bottom boundary condition to
stress the model, with an objective to better simulate the initiation
and evolution of solar explosive phenomena and their interplanetary
evolution process.

\acknowledgments

This work is jointly supported by the 973 program under grant
2012CB825601, the Chinese Academy of Sciences (KZZD-EW-01-4), the
National Natural Science Foundation of China (41204126, 41274192,
41031066, and 41074122), and the Specialized Research Fund for State
Key Laboratories. Data are courtesy of NASA/{SDO} and the HMI science
teams. C.~W. Jiang thanks Dr. X. Luo for a careful revise of the text.
The authors thank the anonymous referee for invaluable comments.



\end{CJK*}
\end{document}